\documentclass[conference]{IEEEtran}
\IEEEoverridecommandlockouts
\usepackage{cite}
\usepackage{amsmath,amssymb,amsfonts}
\usepackage{algorithmic}
\usepackage{graphicx}
\usepackage{textcomp}
\usepackage{xcolor}
\usepackage{makecell}
\usepackage{booktabs} 
\usepackage[hyphens]{url}
\usepackage{svg}
\usepackage{hyperref}
\hypersetup{breaklinks=true}

\newcommand{\lmttfont}{\fontfamily{lmtt}\selectfont}

\def\BibTeX{{\rm B\kern-.05em{\sc i\kern-.025em b}\kern-.08em
    T\kern-.1667em\lower.7ex\hbox{E}\kern-.125emX}}
\begin{document}


\title{Certify the Uncertified: Towards Assessment of Virtualization for Mixed-criticality in the Automotive Domain}

\author{
    \IEEEauthorblockN{Marcello Cinque, Luigi De Simone, Andrea Marchetta}
    \IEEEauthorblockA{Università degli Studi di Napoli Federico II, Naples, Italy
    \\\{macinque, luigi.desimone\}@unina.it}a.marchetta@studenti.unina.it
}

\maketitle

\begin{abstract}
Nowadays, a feature-rich automotive vehicle offers several technologies to assist the driver during his trip and guarantee an amusing infotainment system to the other passengers, too. Consolidating worlds at different criticalities is a welcomed challenge for car manufacturers that have recently tried to leverage virtualization technologies due to reduced maintenance, deployment, and shipping costs. For this reason, more and more mixed-criticality systems are emerging, trying to assure compliance with the ISO 26262 Road Vehicle Safety standard.
In this short paper, we provide a preliminary investigation of the certification capabilities for Jailhouse, a popular open-source partitioning hypervisor. To this aim, we propose a testing methodology and showcase the results, pointing out when the software gets to a faulting state, deviating from its expected behavior. 
The ultimate goal is to picture the right direction for the hypervisor towards a potential certification process.
\end{abstract}


\section{Introduction}

In the last few years, we have witnessed the growing complexity of hardware and software systems for the automotive industry, with the aim of both providing safety and useful driver assisting systems (e.g., parking sensors, cruise control, brake assist, etc.) while having a rich infotainment package with all connectivity features. In this context, the idea of deploying both kinds of subsystems, one with safety-critical functions (often linked to soft/hard real-time constraints) and safety-related standards to comply with and the other with less demanding time constraints in a single platform board is tempting. Indeed, both the integration and development efforts are extremely cut down due to the reduced size, weight, power, and cost of the hardware itself.

To this aim, Mixed-Criticality Systems (MCSs) integrate functionalities of different safety and/or time-critical into a common platform \cite{burns2022mixed}. This kind of system is gradually expanding towards several application fields thanks to the ever-increasing CPU capabilities of modern chips, which have now spread the implementation of MCSs to the embedded domain where power consumption is a major concern. Moreover, the rapid development of virtualization technologies can nowadays easily support mixed-criticality compositions since it implicitly provides software support for resource partitioning and running tasks on heterogeneous Operating System (OS) (real-time and general-purpose) environments \cite{heiser2011virtualizing}. Indeed, due to the recent COVID-19 pandemic, there were estimated losses within ~\$60 to ~\$100 billion in 2021 sales for chip shortages in the automotive industry \cite{bloomberg_chip_shortage}. Thus, virtualization technologies are becoming a prominent way for the industry to consolidate multiple software systems on the same system-on-a-chip (SoC) in a flexible way \cite{blackberry_qnx_chip_shortage, cinque2021virtualizing}.

Generally, MCSs require safety certification evidence, which includes cumbersome and demanding tasks. The problem is even exacerbated when MCSs are virtualized since further critical factors come into play during the process of choosing the proper virtualization solution, which include hypervisor type, memory footprint, certification level (if exists), the license, support to high availability, fault tolerance, and security \cite{cinque2021virtualizing}. Regarding these issues, virtualizing MCSs could be an unfeasible task. However, the reference safety standard in the automotive domain, i.e., the ISO 26262 \cite{iso26262}, allows supplementing an already certified piece of software (e.g., a real-time OS) with a new software element (e.g., a software library) that is not certified, called the \textit{Safety Element out of Context} (SEooC). Including a SEooC in the whole system will not invalidate the former certification level if we can provide proof that this software has been developed following the guidelines specified by ISO 26262, i.e., ensuring that SEooC integration does not negatively affect existing pieces of software. In particular, the standard proposes, among others, to precisely analyze failure modes according to anomalous conditions. In the literature, several studies exploit fault injection techniques to assess complex systems' behavior and unveil potential bottlenecks and critical components under these abnormal inputs and conditions \cite{cotroneo2013fault, cotroneo2018run, nfv_dep_guidelines, winter2015no}. Also, the ISO 26262 standard prescribes the use of error detection/handling mechanisms, and their verification through fault injection, by ``\textit{corrupting hardware or software components}''.

In this short paper, we provide a preliminary investigation on integrating a partitioning hypervisor, namely Jailhouse, in a safety-critical deployment as SEooC. Jailhouse \cite{ramsauer2017look} promises to easily comply with safety or other certification requirements that are difficult to achieve with general-purpose hypervisors \cite{cinque2021virtualizing}.
We decided to focus on fault injection testing, which is an established technique to assess fault tolerance mechanisms via deliberately inject representative faults into a system under test. In our context, we need to provide evidence about isolation guarantees needed for treating a hypervisor as SEooC. To this aim, we propose a preliminary fault injection framework to deliberately inject faults according to the classical bit-flip fault model \cite{natella2016assessing}. Finally, we showcased the results of our initial experiments, highlighting the underlying criticalities of the current version of this hypervisor.

The other sections are organized as follows: section \ref{sec2} provides a general overview of the Jailhouse hypervisor and a discussion about SEooC; section \ref{sec3} describes our testing framework along with preliminary results; section \ref{sec4} delineates the state-of-the-art hypervisors in the automotive industry; section \ref{sec5} concludes the paper.

\section{Jailhouse and ISO 26262}
\label{sec2}

\subsection{Jailhouse Overview}

\begin{figure}
    \centering
    \includegraphics[width=0.83\columnwidth]{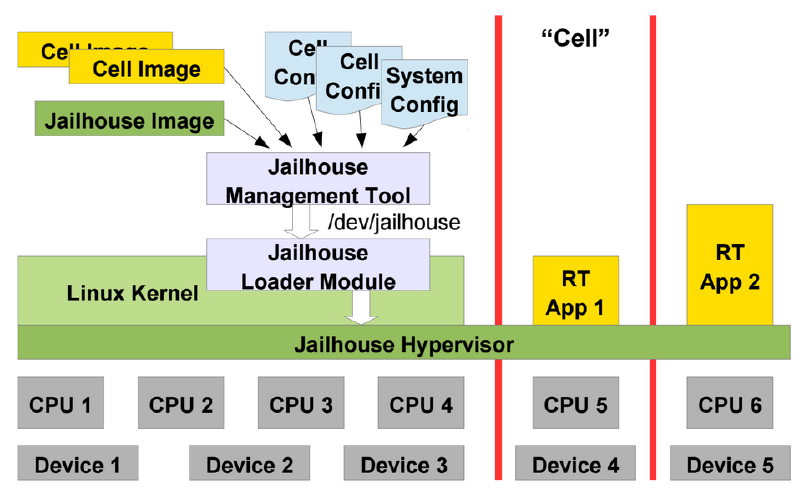}
    \caption{The Jailhouse architecture}
    \label{fig:jailhouse}
\end{figure}

Jailhouse is a hypervisor announced publicly in November 2013 and currently published as open-source under the GNUPLv2 license and available for x86 and ARM-based boards (both 32 and 64 bit) \cite{ramsauer2017look}. Jailhouse is a partitioning hypervisor, which is a type of virtualization where the hardware resources such as cores, peripherals, and memory are divided into partitions (virtual machines) which are allocated statically following the concept of \textit{asymmetric processing}. While this approach may hinder the overall performance, it is desirable in a safety-critical application, as isolation is crucial to ensure no interference for running critical tasks. Further, a partitioning solution also keeps the overall codebase small, making easier (theoretically) the certification process. In Jailhouse, every partition is called \textit{"cell"}. As previously said, every cell has its own resources which cannot be accessed by the others. When Jailhouse is started, the only cell placed in the system is the \textit{root cell}. This cell contains every resource that Jailhouse can share between all the other cells, and thus it can not be removed from the system. Jailhouse allows creating a static configuration for a cell by writing a source file according to special C structures, where each field is filled according to the customer needs (e.g, assigned CPU cores, memory areas and access permissions, IRQ enabled, etc.).
Generally, Jailhouse statically assigns to each cell one guest OS (optional for bare-metal applications) and its applications called \textit{inmates}. Despite the main objective being partitioning resources, inter-cell communication is allowed through the \textit{ivshmem} device model. \figureautorefname{}~\ref{fig:jailhouse} shows the architecture of Jailhouse.

\subsection{Jailhouse as Safety Element out of Context (SEooC)}

Jailhouse is currently still in beta, as the latest version is the v0.12 on February 2020. However, this hypervisor was recently used as building block in different European research projects \cite{selene_project, hercules2020} and academic studies.
In this short paper, our aim is to understand whether this kind of software can be leveraged as a Safety Element out of Context, as described in the ISO 26262 road safety standard. Citing this latter, \textit{"An SEooC can be [...] a software [which is] developed, based on assumptions, in accordance with ISO 26262. It is intended to be used in multiple different items when the validity of its assumptions can be established during integration of the SEooC."}
Jailhouse can fall into this category: while it has not been developed as a vehicle software component, it can still be certified as such if its development is considered compliant to the ISO 26262 standard, and if it is not, it can be suitable for certification with change. Unfortunately, at least in public documents, there is no official documentation which describes the software planning activity nor the design phase of Jailhouse, thus this whole documentation should be compiled from scratch. Moreover, following the ISO 26262 standard, the software has to be tested with a software component which is certified as ISO 26262 with a certain ASIL (Automotive Safety Integrity Level), and in case the safety assumptions can not fit the new context, an impact analysis has to be initiated.
Ultimately, the process of certifying Jailhouse as a SEooC would be far from near completion. However, granting that both root and non-root cells can't interfere with each other, thus avoiding any issues with potential safety-related tasks, is a starting point for it in order to leverage the application to more safety concerning industrial applications.


\section{Proposal and Preliminary Results}
\label{sec3}

\begin{figure}
    \centering
    \includegraphics[width=0.8\columnwidth]{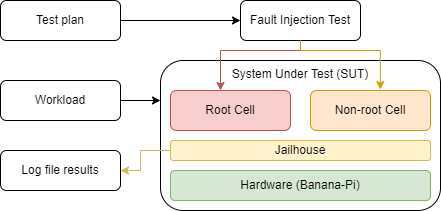}
    \vspace{-0.3cm}
    \caption{Testing Framework}
    \label{fig:test-framework}
    \vspace{-0.7cm}
\end{figure}

Our preliminary work aims at verifying the Jailhouse isolation and integrity features regarding the ability to provide cell partitioning. \figureautorefname{}~\ref{fig:test-framework} shows the proposed testing framework: after we decided on a certain test plan, we settled for a fault injection test with a certain hardware setup that affected both the root and non-root cell deployed by the hypervisor. Results were therefore collected on a log file, used further for analytics. The proposed testing framework includes a dozen of lines of code added to Jailhouse that allows us to orchestrate the fault injection tests by controlling test duration and target. We leveraged the single bit-flip fault model commonly used to emulate transient hardware faults \cite{natella2016assessing}. Having the code source at our disposal, we decided to monitor some golden (fault-free) runs of the hypervisor in order to find preliminary fault injection points. This profiling operation yielded three candidates functions, which were the hardware interrupt request function ({\lmttfont irqchip\_handle\_irq()}), the trap exception handler ({\lmttfont arch\_handle\_trap()}) and the hypervisor call handler ({\lmttfont arch\_handle\_hvc()}). These three functions refer to the virtualization extensions available for the ARMv7 architecture. We then chose the injection strategy, consisting of a random bit flipping of a random architecture register. Finally, we decided not to operate on the interrupt request handler as the only parameter passed is the IRQ vector number, and manumitting it means calling a different IRQ function, defaulting to an IRQ error, which is completely predictable and correct behavior.

The generated test plan consists of two classes of testing, defined by the fault intensity level: the \textit{medium level} refers to a discontinuous bit flipping of a single register, generated once every given number of calls to the target functions, while the \textit{high level} instead consists in a bit flip of multiple registers at the time. These are both transient fault injections, and the rate of occurrence is configurable. The showcased tests have an occurrence of once every $100$ and $50$ function calls for the medium and hard intensity, respectively.
Each test lasts 1 min. and the outcome is sent to an empty shell where the board serial port is connected. We then collected the test results into a log file, which is further analyzed to understand how the hypervisor reacted to injected faults. The tested hardware comprises a Banana PI, which is a dual-core Cortex-A7 board, equipped with 1 GB of RAM. We evaluated Jailhouse v0.12 with Linux Kernel v5.10, properly patched for Jailhouse.
Finally, the test plan was executed by exercising a workload consisting of a root cell where the general-purpose Linux was running and a non-root cell in which we run FreeRTOS, a market-leading real-time OS. In particular, within FreeRTOS we spawned several tasks to be managed, including a task to blink an onboard led, a couple of send/receive tasks, two floating-point arithmetic tasks, and fifteen integer ones. We statically assigned the board CPU core 0 to the root cell and the CPU core 1 to the non-root cell (FreeRTOS cell). 

The results can be summarized as follows. \textit{High level} intensity faults always return an "invalid arguments" when we target both the {\lmttfont arch\_handle\_hvc()} and {\lmttfont arch\_handle\_trap()} in the context of the root cell; thus, the root cell will be not allocated at all, which is a correct (and expected) behavior. On the other hand, when we filter the injection to activate only when the CPU core 1 is calling the function, the result is pretty peculiar, although wrong and inconsistent: the cell is allocated but, whether the CPU fails to come online as per the swap feature of the CPU hot plug or the cell is left in a non-executable state, the non-root cell doesn't do anything, as attested by the USART output left completely blank. Nonetheless, it is considered \textit{running} by Jailhouse, and the shutdown of the cell (i.e., {\lmttfont jailhouse cell shutdown} command) gives the control of the CPU and the non-root cell peripherals specified in the configuration file back to the root cell. This inconsistent state is particularly dangerous as the Jailhouse user assumed that the allocated non-root cell is running, but instead, it is completely broken and unusable, and only destroying the cell and reallocating it fixes the problem.
We then decided to further investigate the non-root cell {\lmttfont arch\_handle\_trap()} call to see the extend of the inconsistency, yielding the results summarized in \figureautorefname{}~\ref{fig:non-root-availability-measure}:

\begin{figure}[h]
    \centering
    \includegraphics[width=0.7\columnwidth]{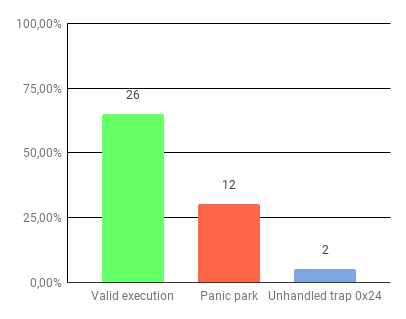}
    \vspace{-0.3cm}
    \caption{Non-root cell availability in medium intensity tests}
    \label{fig:non-root-availability-measure}
\end{figure}

As we can see, the cell behaves correctly in the majority of cases, although in the $30\%$ a \textit{panic park} happens, i.e., the fault propagates to the whole system bringing the system itself to a kernel panic. Finally, a limited number of tests brings to a \textit{CPU park}: while the non-root cell is running, there is a critical fault injection that triggers the error code {\lmttfont 0x24}, which is the unhandled trap exception. In this case, the {\lmttfont cpu\_park()} function is called and the non-root cell stops working. However, the destruction of the non-root cell, which brings the CPU core 1 control back to the root cell, is accomplished without any issue. This means that the fault has been successfully isolated and the non-root cell has not damaged the other cells.

\section{Related Work}
\label{sec4}

\textit{PikeOS} \cite{pikeos} is a commercial hard real-time operating system that offers a separation kernel-based hypervisor. This real-time OS (RTOS) allows the creation of logical partitions which are completely isolated and can execute with different safety and security levels within the same computing platform. PikeOS is developed according to various safety standards, like DO-178C for avionic and also ISO 26262. The main advantage of the PikeOS architecture is that the microkernel consists of real-time OS and a type-1 hypervisor in one product. The system achieves resource partitioning by statically assigning the computing resources to its partitions and by using MMU.
\textit{MICROSAR} \cite{microsar} includes in its OS package a preemptive, real-time multitasking operating system for microcontrollers. It is based on the AUTOSAR OS specification \cite{autosar}, which is an extension of the OSEK/VDX-OS standard \cite{osek_vdx}, offering functions for time monitoring and memory protection. It also implements a high-resolution timer mechanism in order to provide time resolutions of less than $1ms$. The OS package is provided by a hypervisor module called LeanHypervisor (vLhyp) which initializes the System Memory Protection Unit (MPU) at start-up and manages the CPU cores' power on. Each core may have its own operating system image, which can be POSIX, Classic, or Adaptive AUTOSAR. Both the hypervisor and the Memory Protection Unit (MPU) task protect the OS partitions, which have to run without the risk of mutual interference due to incorrect data changes so that the system can operate in parallel partitions with different ASILs.
Other than the OS and hypervisor modules, there are plenty of other modules which compose the MICROSAR Safe basic software which have all been certified by the ISO 26262 ASIL-D standard as SEooC \cite{autosar-certified}.
\textit{VOSYSmonitor} \cite{lucas2018vosysmonitor} is an ASIL-C ISO 26262-compliant hypervisor, developed for both ARMv7 and ARMv8 architectures whose Cortex-A processors offer the \textit{Secure Monitor} mode. This feature allows executing a safety-critical RTOS and a general-purpose OS simultaneously; moreover, it allows the use of virtualization extensions like Linux/KVM, thus enabling the instantiation of multiple virtual machines. Furthermore, the hypervisor isolates the RTOS from the other VMs and offers several communication methods in order to provide a secure communication channel between the various subsystems. On multicore architectures, VOSYSmonitor can dynamically share cores between the \textit{secure} side and the \textit{non-secure} one: the idea is that the \textit{secure world} tasks should necessarily have priority compared to the other ones. This means that once a core is assigned to the safety-critical domain, the \textit{non-secure world} applications can run on that core only if there are no critical tasks to run, thus meaning that the resource has been freed by the \textit{secure-world} application. An interruption mechanism is finally deployed in order to guarantee an efficient context switch from one world to another.
Besides commercial products, also in research there is a trend toward the use of static partitioning hypervisors, mainly due to the isolation property that this kind of virtualization can offer to the user. 
We choose Jailhouse as our testing platform due to its adaptability to multiple boards and the fact that it is open-source, but there are other successful examples of commercial hypervisors for automotive platforms currently available on the market. Other than Jailhouse, another example is \textit{Bao} hypervisor \cite{martins2020bao}, proposed as a static partitioning hypervisor for Armv8 and RISC-V platforms with a small codebase which differs from Jailhouse due to the fact that it does not depend on Linux to boot and manage the partitions. To the best of our knowledge, this is the first study that investigates fault injection techniques to support the certification of a hypervisor as SEooC in the automotive domain.

\section{Conclusions and future work}
\label{sec5}
In this short paper, we investigated the possibility of treating the open-source, non-certified, Jailhouse partitioning hypervisor as a Safety Element out of Context (SEooC) according to the automotive reference standard, the ISO 26262. More specifically, we opted for a fault injection methodology aimed at verifying the isolation and integrity features provided by the Jailhouse hypervisor. At the best of our knowledge, there are no other researches that tackle a fault injection approach towards verifying the hypervisor robustness and isolation features. In the end, we highlighted some possible criticalities which led to the hypervisor malfunctioning. Future plans aim at expanding the fault injection testing framework, by applying, e.g., a wider and customizable set of fault models in real scenarios. This will enable a framework that significantly helps this kind of hypervisor to become actual building blocks in virtualized mixed-criticality systems.


\bibliographystyle{IEEEtran}
\bibliography{bibliography}

\end{document}